\documentclass[a4paper]{jpconf}

\usepackage{subfigure}
\usepackage{graphicx}
\usepackage{amssymb}
\usepackage{amsmath}
\usepackage{amstext}
\usepackage{braket}
\usepackage[english]{babel}
\usepackage[latin1]{inputenc}

\newcommand{\vek}[1]{\ensuremath{\textbf{#1}}}

\begin{document}
\title{Similarity of eigenstates in generalized labyrinth tilings}

\author{Stefanie Thiem and Michael Schreiber}
\address{Institut f\"ur Physik, Technische Universit\"at Chemnitz, D-09107 Chemnitz, Germany}
\ead{\{stefanie.thiem, michael.schreiber\}@physik.tu-chemnitz.de}

\begin{abstract}
The eigenstates of $d$-dimensional quasicrystalline models with a separable Hamiltonian are studied within the tight-binding model. The approach is based on mathematical sequences, constructed by an inflation rule $\mathcal{P} = \{w \rightarrow s, s \rightarrow sws^{b-1}\}$ describing the weak/strong couplings of atoms in a quasiperiodic chain. Higher-dimensional quasiperiodic tilings are constructed as a direct product of these chains and their eigenstates can be directly calculated by multiplying the energies $E$ or wave functions $\Psi$ of the chain, respectively.

Applying this construction rule, the grid in $d$ dimensions splits into $2^{d-1}$ different tilings, for which we investigated the characteristics of the wave functions. For the standard two-dimensional labyrinth tiling constructed from the octonacci sequence ($b=2$) the lattice breaks up into two identical lattices, which consequently yield the same eigenstates. While this is not the case for $b \ne 2$, our numerical results show that the wave functions of the different grids become increasingly similar for large system sizes. This can be explained by the fact that the structure of the $2^{d-1}$ grids mainly differs at the boundaries and thus for large systems the eigenstates approach each other. This property allows us to analytically derive properties of the higher-dimensional generalized labyrinth tilings from the one-dimensional results. In particular participation numbers and corresponding scaling exponents have been determined.
\end{abstract}

\section{Introduction}

Since the discovery of quasicrystals research aims at understanding the physical properties of these materials. Today many exact results are known for one-dimensional quasicrystals \cite{JPhysFrance.1989.Sire, MathQuasi.2000.Damanik} but the characteristics of wave functions in two or three dimensions have been clarified to much lesser degree. In particular, they have mainly been studied by numerical calculations of finite systems or periodic approximants due to the missing translational symmetry of these materials. For instance numerical results for the Penrose tiling, the Ammann-Kramer-Neri tiling or octagonal tiling models have been determined but for relatively small systems making it difficult to estimate the behavior for macroscopic system sizes \cite{Quasicrystals.2003.Grimm, PhysRevB.1992.Passaro}. Further, it is possible to calculate some exact eigenstates of the tight-binding Hamiltonian on the Penrose tiling \cite{PhysRevB.1995.Rieth, PhysRevB.1998.Repetowicz}, which is a first step to understand the nature of eigenstates and properties of quasicrystals in higher dimensions.

In order to get a better insight into the properties of wave functions in quasicrystals, another approach employs models of
$d$-dimensional quasicrystals, which can be built from $d$ separate one-dimensional systems yielding straightforward solutions. This
allows us to study very large systems in higher dimensions based on the solutions in one dimension \cite{Ferro.2004.Ilan}. Unfortunately, these structures do not occur in real quasicrystals, however, they can be artificially constructed. Examples are conducting
nanowires, coupled nanomechanical resonators, or photonic quasicrystals \cite{Ferro.2004.Ilan, PhysWorld.2004.McGrath}. The best studied
model within this approach is based on the Fibonacci sequence corresponding to the golden mean \cite{JPhysFrance.1989.Sire,Ferro.2004.Ilan}, whereas in this paper we concentrate on the other metallic mean quasiperiodic sequences
\cite{JPhysFrance.1989.Sire2, PhysRevB.2000.Yuan, JPhysA.1989.Gumbs}, which have been less investigated so far.

At first we introduce the quasicrystal model based on metallic mean sequences and discuss the symmetry properties of the eigenstates in Sec.~\ref{sec:chains}. This approach is expanded to higher dimensions for the labyrinth tiling in Sec.~\ref{sec:labyrinth}. In general, the grid splits into $2^{d-1}$ different tilings in $d$ dimensions but the results indicate that the wave functions of these tilings approach each other for large systems. We give a reasonable argument for this behavior as well. This property can be used to calculate or even analytically derive characteristics for generalized labyrinth tilings. As an example, Sec.~\ref{sec:participation} comprises results for the participation ratio of the eigenstates and their scaling behavior in one, two, and three dimensions. The paper is concluded in Sec. \ref{sec:conclusion}.

\section{Eigenstates and wave functions of metallic mean chains}\label{sec:chains}

The metallic mean sequences for a parameter $b$ are defined by the inflation rule $\mathcal{P} = \{w \rightarrow s, s \rightarrow sws^{b-1}\}$
based on an alphabet containing the symbols $\{s,w\}$ with the starting symbol $w$.
After $a$ iterations we obtain the $a$th order approximant $\mathcal{C}_a$ of the quasiperiodic chain. Regarding the length $f_a$ of this approximant $\mathcal{C}_a$ we can also find a recursive rule $f_a = b f_{a-1} + f_{a-2}$ with $f_0 = f_1 = 1$. The ratio of the lengths of two successive iterants in the limit $ \lim_{a \rightarrow \infty} f_a / f_{a-1} = \lambda$ is given by different metallic means depending on $b$ with a continued fraction representation $\lambda = [\bar{b}] = [b,b,b,...]$. This leads to the well known Fibonacci sequence for $b=1$ with the golden mean $\lambda_{\mathrm{Au}} = [\bar{1}]=(1+\sqrt{5})/2$, while $b=2$ results in the octonacci sequence with silver mean $\lambda_{\mathrm{Ag}} = [\bar{2}] = 1+\sqrt{2}$ and $b=3$ corresponds to the bronze mean $\lambda_{\mathrm{Bz}} = [\bar{3}] = (3+\sqrt{13})/2$ \cite{JPhysA.1989.Gumbs}.

For the orthogonal basis states $\ket{l}$ the tight-binding Hamiltonian can be written as
\begin{equation}
 \label{equ:octonacci.2}
 \boldsymbol{\mathcal{H}} = \sum_{l=1}^{f_a} \ket{l} t_{l,l+1} \bra{l+1} + \sum_{l=1}^{f_a+1} \ket{l}\varepsilon_l\bra{l}  \;.\\
\end{equation}
This can be interpreted as an electron on a discrete grid which can hop from one vertex of this graph to any of its neighboring vertices and the aperiodicity of the grid is given by the quasiperiodic sequence with the symbols $w$ and $s$ corresponding to weak and strong bonds. Hence, the hopping parameter $t$ is determined according to the sequence $\mathcal{C}_a$ with $t_{s} = 1$ for a strong bond and $t_{w} = v$ for a weak bond ($0 \le v \le 1$). Further, usually one assumes zero on-site potentials ($\varepsilon_l = 0$) for these models so that no vertex is energetically preferred \cite{Quasicrystals.2003.Grimm, PhysRevB.2000.Yuan}.

The discrete energy values $E^i$ and corresponding wave functions $\ket{\Psi^i} = \sum_{l=1}^{f_a+1} \Psi_l^i \ket{l}$ are solutions of the time-independent Schrödinger equation for the quasiperiodic systems
\begin{equation}
 \label{equ:octonacci.8}
 \mathcal{H} \ket{\Psi^i} = E^i \ket{\Psi^i} \Longrightarrow E^i \Psi_l^i = t_{l-1,l} \Psi_{l-1}^i + t_{l,l+1} \Psi_{l+1}^{i} \;.
 \end{equation}

The results show that the eigenvalues are symmetric with respect to $0$. However, we have to distinguish between even and odd system sizes $N_a = f_a + 1$: for even $N_a$ all energy values $E$ have a symmetric value $-E$ but for odd $N_a$ there is one state $E^M = 0$, which has no corresponding state.
Additionally, also the eigenfunctions show a symmetry. The eigenstates $\Psi$ with an eigenvalue $E$ and $\widetilde{\Psi}$ with the
opposite eigenvalue $-E$ only differ by an alternating sign depending on the position $l$ with
 \begin{equation}
  \label{equ:octonacci.6}
  \widetilde{\Psi}_l = (-1)^l \Psi_l \;.
 \end{equation}
Further, for odd $N_a = 2M-1$ the eigenvector $\Psi_l^M$ associated to the eigenvalue $E^M = 0$ has a special structure, namely $\Psi_l^M$ vanishes either on all odd or on all even sites $l$ (cp. \cite{PhysRevB.2000.Yuan}), i.e.,
 \begin{align}
  \label{equ:octonacci.7}
   \Psi_{l\bmod 2 \equiv 0}^{M^-} &= 0 & \text{or}& &   \Psi_{l\bmod 2 \equiv 1}^{M^+} &= 0 \;.
 \end{align}

\section{Eigenstates and wave functions of the generalized labyrinth tiling}\label{sec:labyrinth}

\begin{figure}[b]
 \centering
  \subfigure[silver mean model $\mathcal{L}_3^{\textrm{Ag}\star}$]{\includegraphics[width=5.0cm]{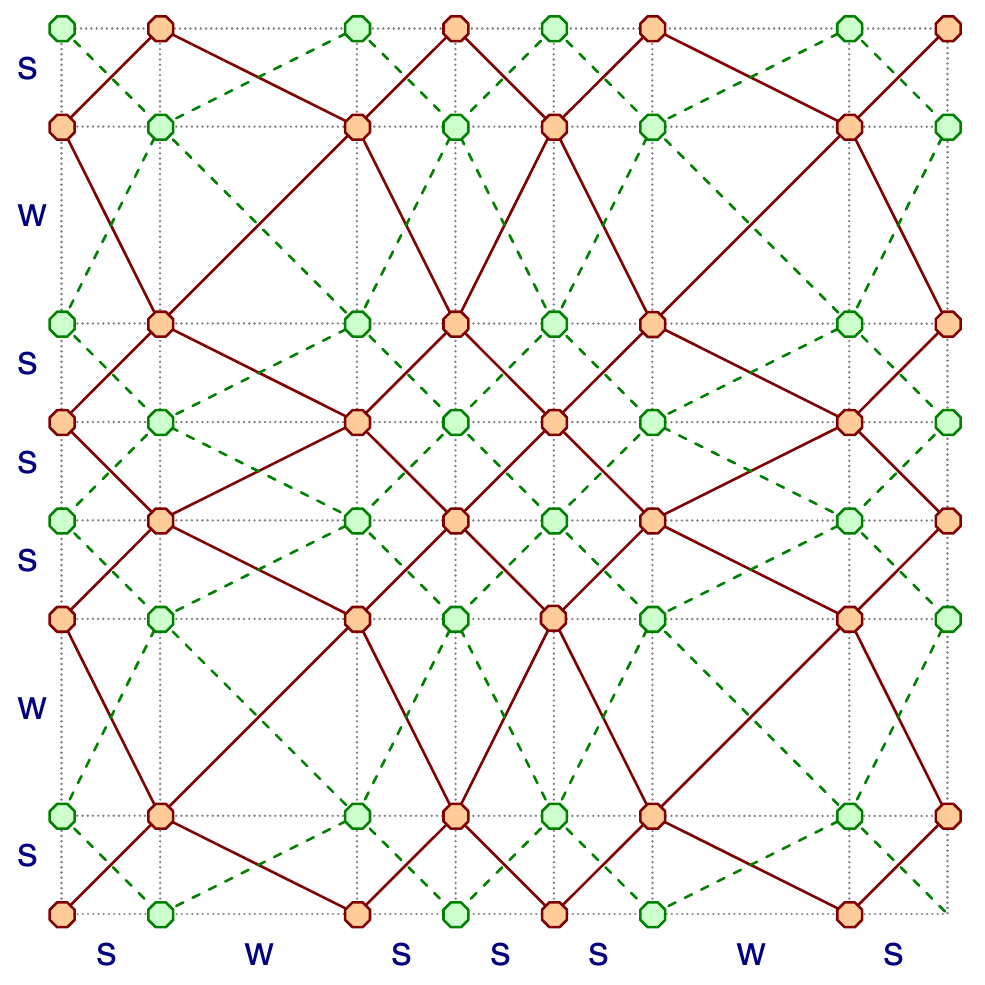}} \hspace{0.2cm}
  \subfigure[golden mean model $\mathcal{L}_7^{\textrm{Au}}$]{\includegraphics[width=5.0cm]{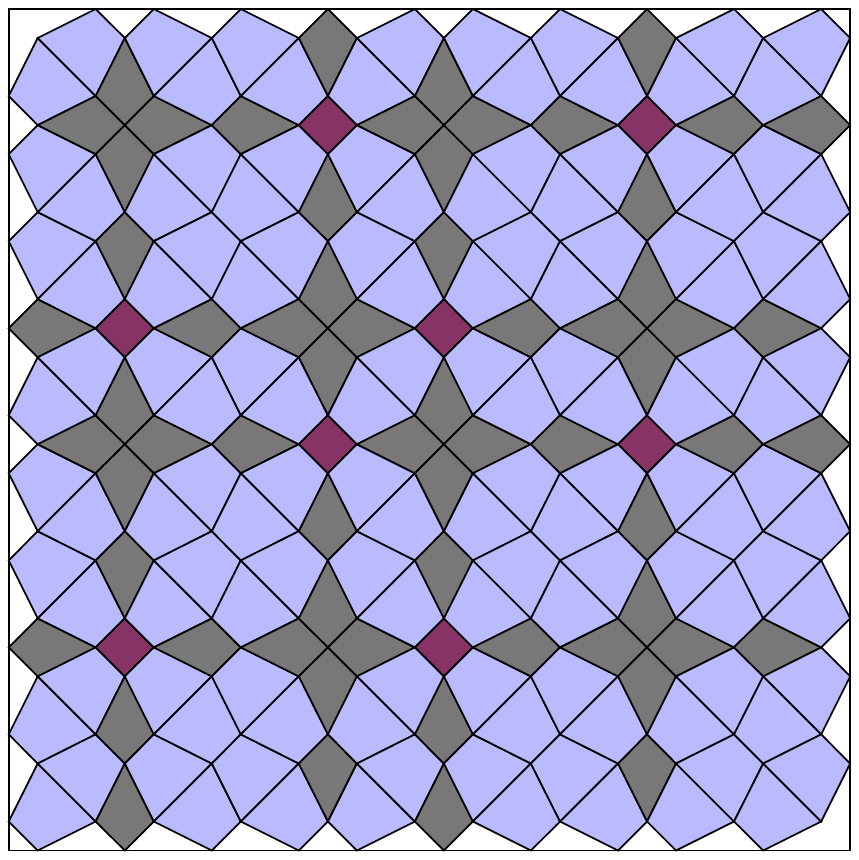}} \hspace{0.2cm}
  \subfigure[golden mean model $\mathcal{L}_7^{\textrm{Au}\star}$]{\includegraphics[width=5.0cm]{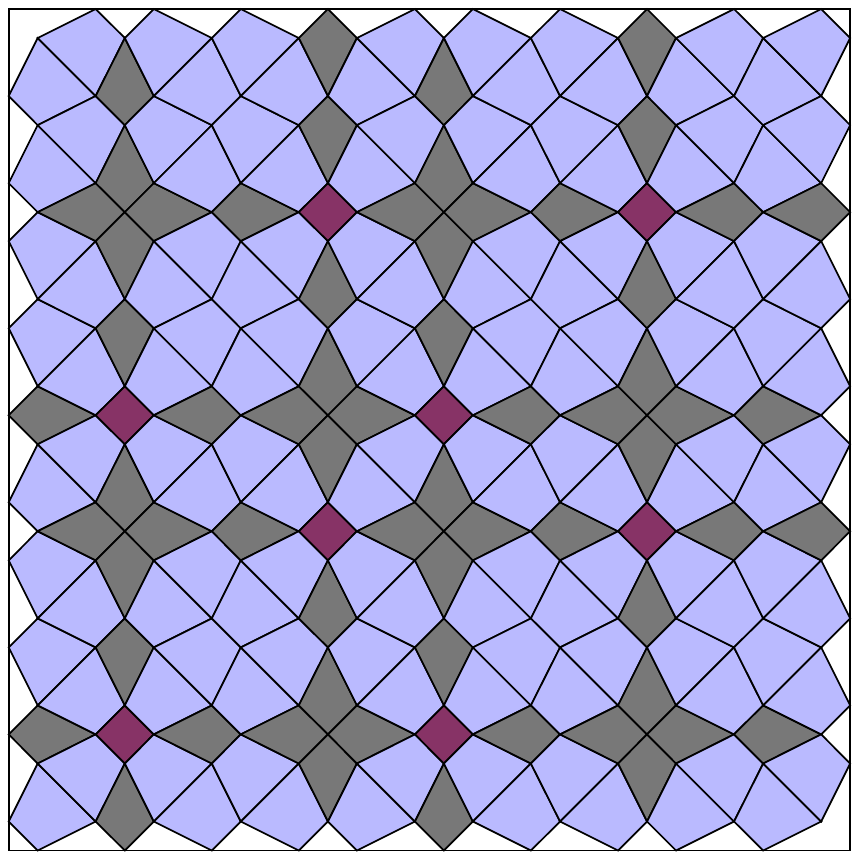}} \vspace{-0.1cm}
 \caption{Labyrinth tilings $\mathcal{L}_a$ and $\mathcal{L}_a^\star$ for different inflation rules: (a) visualizes the construction of $\mathcal{L}_a$ and $\mathcal{L}_a^\star$ for the octonacci chain; (b) shows $\mathcal{L}_7^{\textrm{Au}}$ and (c) $\mathcal{L}_7^{\textrm{Au}\star}$ for the golden mean model, where $\mathcal{L}_a$ is mirrored at the centre of the y-axis to show the similarities of both grids.}
 \label{fig:2Dlattice2}
\end{figure}

The generalized labyrinth tiling is based on two quasiperiodic chains perpendicular to each other. Hence, the labyrinth tiling can be expressed as the Euclidean product $\mathcal{L}_a =  C_a \times C_a$ of the two chains. The bonds are given by the diagonals of the constructed grid and depending on the starting point the grid decomposes into two separate ones, cp.  Fig.~\ref{fig:2Dlattice2}(a). Connecting all vertices with even index sum, we obtain the finite $a$th order approximant $\mathcal{L}_a$ of the generalized labyrinth tiling $\mathcal{L}$. Analogously, all vertices with odd index sum connected by diagonal bonds also form a labyrinth tiling $\mathcal{L}_a^\star$ \cite{JPhysFrance.1989.Sire2, PhysRevB.2000.Yuan}. Both tilings are shown for different inflation rules in Fig.~\ref{fig:2Dlattice2}. For the octonacci sequence they are identical if $\mathcal{L}_a$ is mirrored at the centre of the x- or y-axis or rotated by 90$^\circ$, i.e.~$\mathcal{L}_a^\star$ is dual to $\mathcal{L}_a$. For other inflation rules $\mathcal{L}_a$ and $\mathcal{L}_a^\star$ are slightly shifted against each other and mainly differ at the boundaries as shown in Fig.~\ref{fig:2Dlattice2}(b) and \ref{fig:2Dlattice2}(c). These similarities originate from hidden mirror symmetries in the quasiperiodic chains. While the octonacci chain is perfectly mirror symmetric, for the other chains mirror symmetry can be achieved by neglecting a few symbols at one end of the sequence (e.g. for the Fibonacci chain the last two symbols on the right) and by interchanging two consecutive bonds for the case $b>2$ (i.e.~a phason flip).

The eigenstates of the generalized labyrinth tiling are constructed as the product of the eigenstates of two one-dimensional chains, i.e. $E^{ij} = E^{1i} E^{2j}$ and  $\Phi_{lm}^{ij} \sim \Psi_{l}^{1i} \Psi_{m}^{2j}$ \cite{PhysRevB.2000.Yuan}. The indices $i$ and $j$ enumerate the eigenvalues $E$ in ascending order and the corresponding wave functions $\Psi$, and $l$ and $m$ represent the coordinates of the vertices. However, due to the symmetries of the eigenfunctions of Eq.~\eqref{equ:octonacci.6}, some of the wave functions $\Phi_{lm}^{ij}$ and related eigenvalues $E^{ij}$ are identical. Then only one of them is allowed to be considered. In two dimensions all combinations
 \begin{equation}
  \label{equ:labyrinth.4a}
  \left\{ E^{ij} \,|\, 1\le i \le N_a/2 \wedge 1 \le j \le N_a \right\}
 \end{equation}
are valid for even chain lengths $N_a$. Further, the wave functions $\Phi_{lm}^{ij}$ have to be normalized. For odd $N_a$ the selection of the  eigenstates is more complicated due to the special structure of the eigenfunction for $E^M = 0$. This results in the two-dimensional energy values
 \begin{equation}
  \label{equ:labyrinth.4b}
  \left\{ E^{ij} \,|\, \left(1\le i< M \wedge 1 \le j \le N_a \right) \vee \left( i=M \wedge 1 \le j \le M \right) \right\} \;.
 \end{equation}

In three dimensions we follow the same considerations as in two dimensions. The grid now decomposes into four different grids depending on
the choice of the starting point because the bonds correspond to one of the four main diagonals of a cuboid.
Without loss of generality we consider only the labyrinth tiling $\mathcal{L}_a^{3\text{d}} = \mathcal{C}_a \times \mathcal{C}_a \times \mathcal{C}_a$, for which all vertices have either all even or all odd indices $l$, $m$, and $n$.

We analogously compute the eigenvalues and vectors in three dimensions. Here they are based on three independent quasiperiodic chains, where each one is perpendicular to the other two chains. The wave functions are now given as a product of three one-dimensional eigenfunctions with $\Phi_{lmn}^{ijk} \sim \Psi_{l}^{1i}  \Psi_{m}^{2j} \Psi_{n}^{3k}$, where $i$, $j$, and $k$ label the energy values and $l$, $m$, and $n$ the vertices of the tiling. For even $N_a$ the eigenvalues $E^{ijk} = E^i E^j E^k$ of the generalized labyrinth tiling $\mathcal{L}_a^{3\text{d}}$ in three dimensions are given by the set
 \begin{equation}
  \label{equ:labyrinth3D.3a}
   \left\{ E^{ijk} \,|\, 1\le i \le N_a/2 \wedge 1 \le j \le N_a/2 \wedge 1 \le k \le N_a \right\} \;.
 \end{equation}
This equation results from the symmetry of the wave functions from Eq.~\eqref{equ:octonacci.6}. Due to the structure of the eigenfunctions to the eigenvalue $E^M = 0$, for odd $N_a$ the allowed combinations are given by
  \begin{multline}
    \label{equ:labyrinth3D.3b}
   \{ E^{ijk} \,|\, \left(1\le i < M \wedge 1 \le j < M \wedge 1\le k \le N_a \right) \vee \\
                     \left(i = M \wedge 1 \le j \le M \wedge 1 \le k \le M \right)  \vee
                    \left(1 \le i < M \wedge j = M \wedge 1 \le k \le M \right)  \} \;.
  \end{multline}

\begin{figure}[b]
 \centering
  \subfigure[$\Phi$ for $\mathcal{L}_6^{\textrm{Ag}}$ ($E = 0.4640$)] {\includegraphics[width=4.2cm]{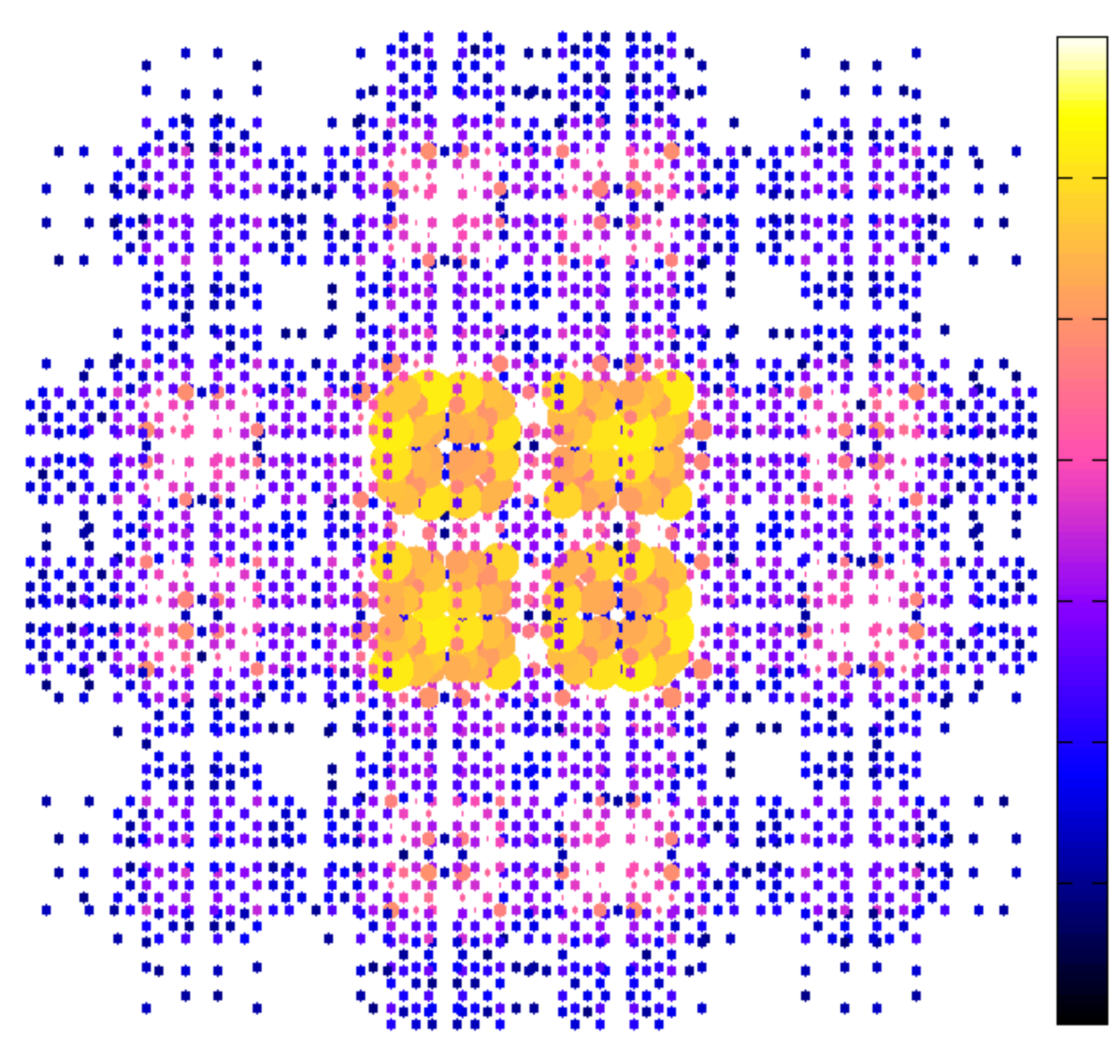}}\hspace{0.5cm}
  \subfigure[$\Phi$ for $\mathcal{L}_{10}^{\textrm{Au}}$ ($E = 0.9239$)] {\includegraphics[width=4.2cm]{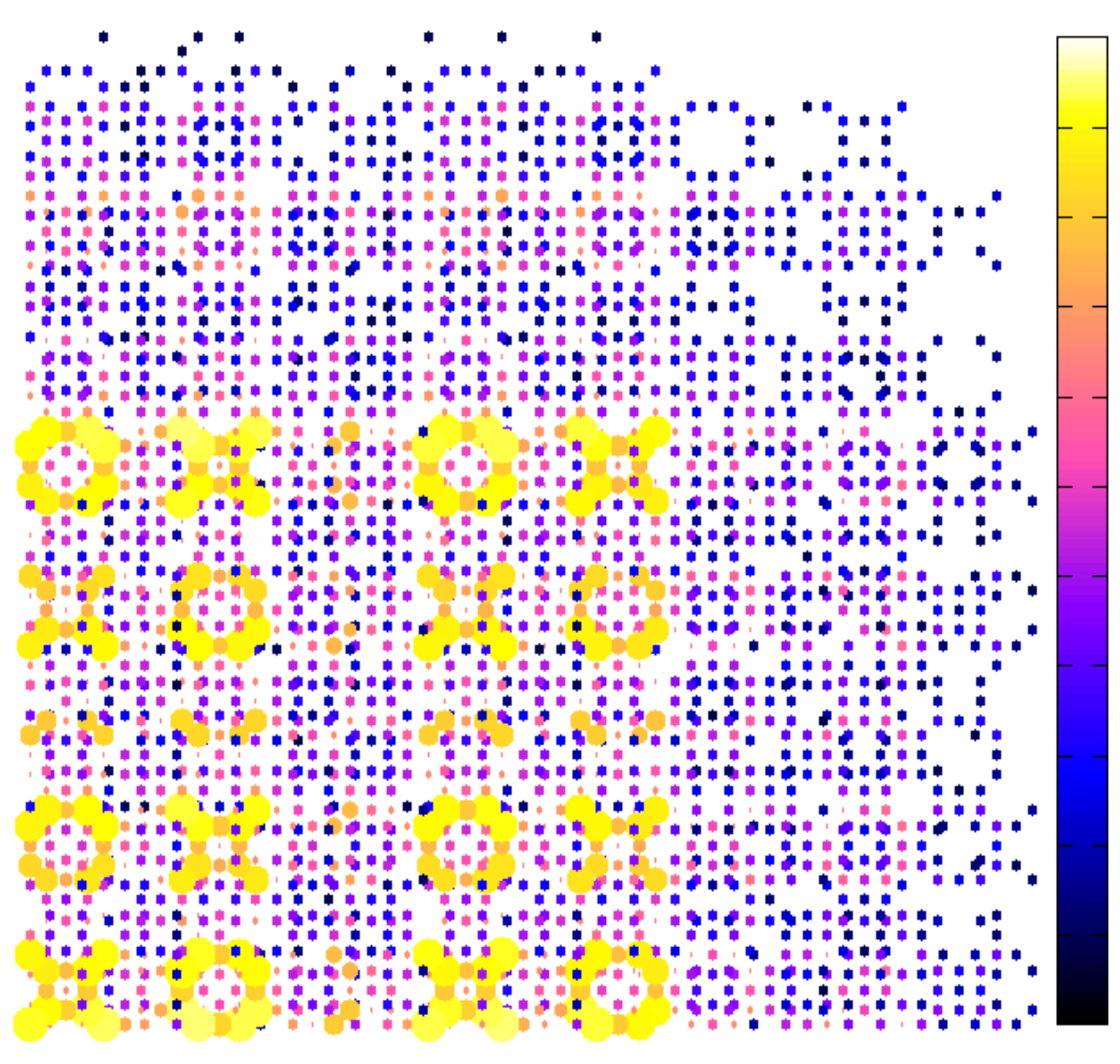}}\hspace{0.5cm}
  \subfigure[$\Phi$ for $\mathcal{L}_{15}^{\textrm{Au}}$ ($E = 2.3518$)]
  {\includegraphics[width=4.2cm]{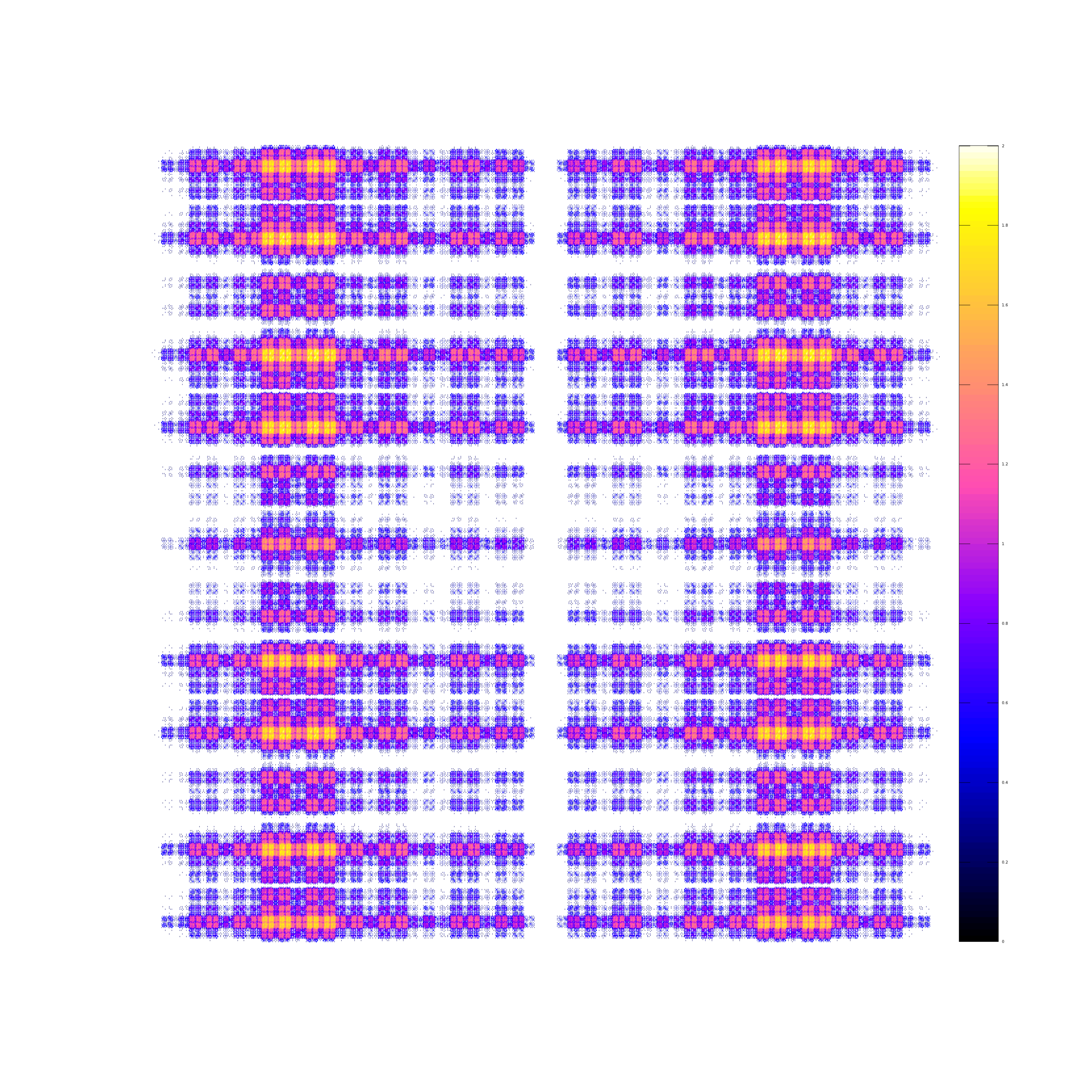}}\vspace{-0.1cm}
  \subfigure[$\Phi$ for $\mathcal{L}_6^{\textrm{Ag}\star}$ ($E = 0.4640$)] {\includegraphics[width=4.2cm]{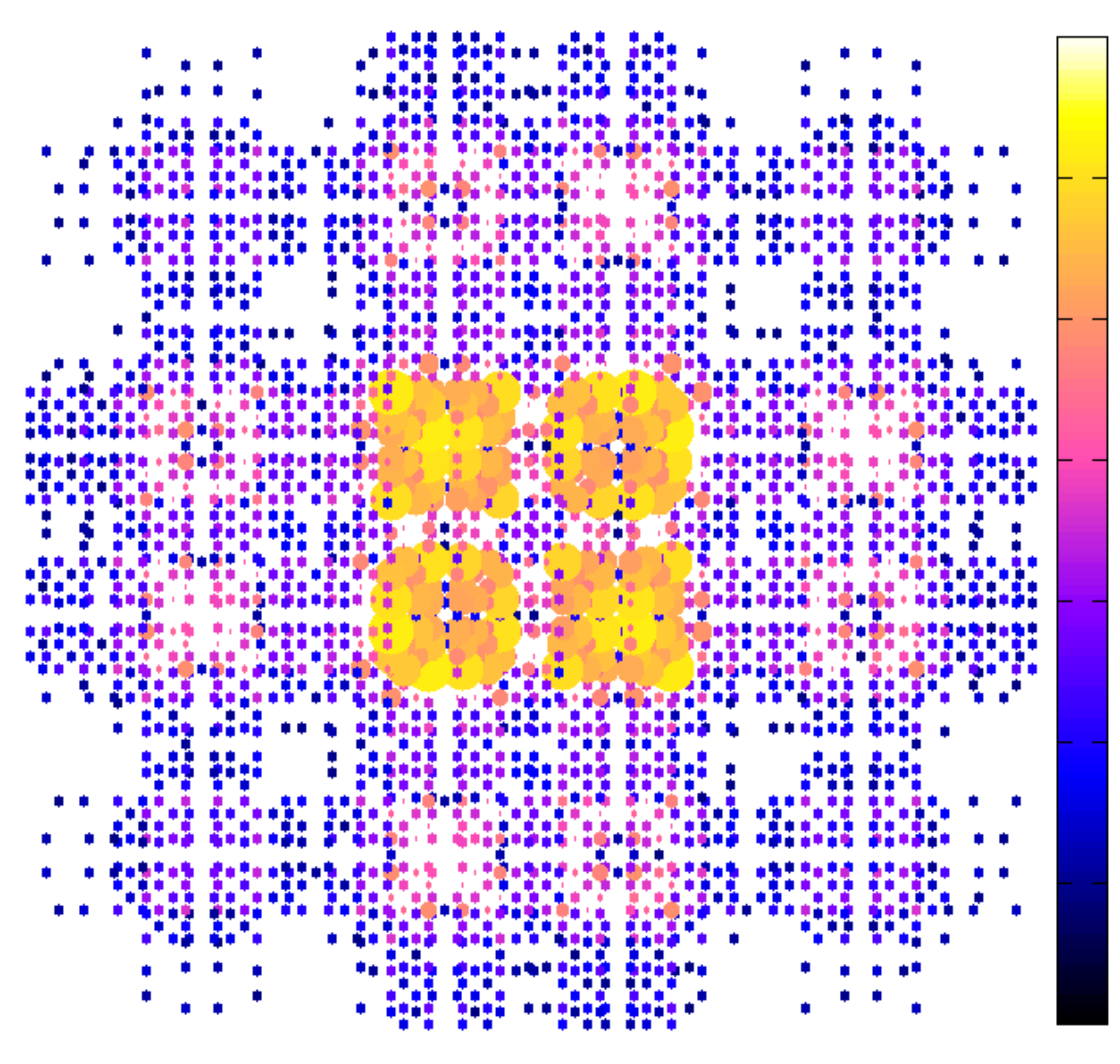}}\hspace{0.5cm}
  \subfigure[$\Phi$ for $\mathcal{L}_{10}^{\textrm{Au}\star}$ ($E = 0.9239$)] {\includegraphics[width=4.2cm]{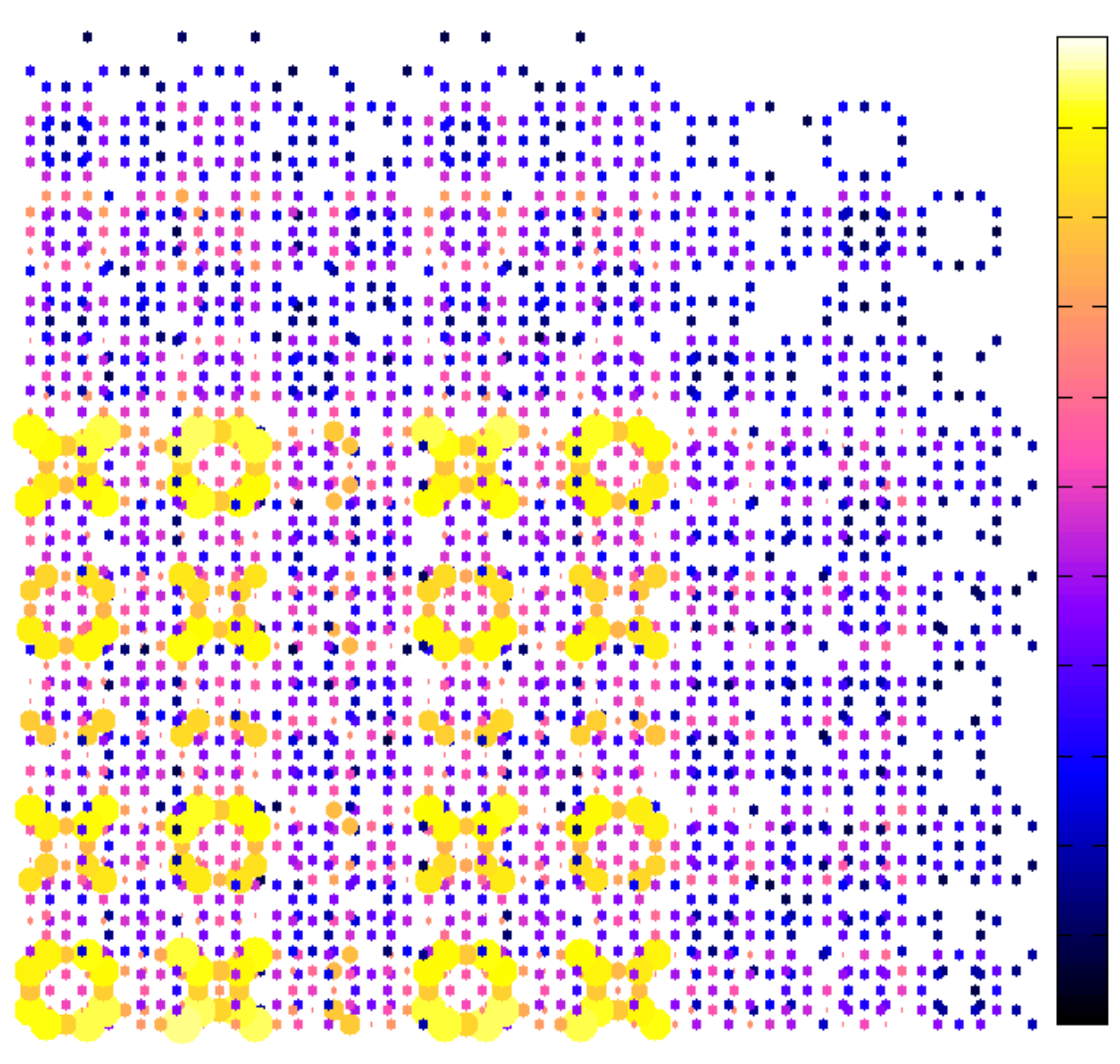}}\hspace{0.5cm}
  \subfigure[$\Phi$ for $\mathcal{L}_{15}^{\textrm{Au}\star}$ ($E = 2.3518$)] {\includegraphics[width=4.2cm]{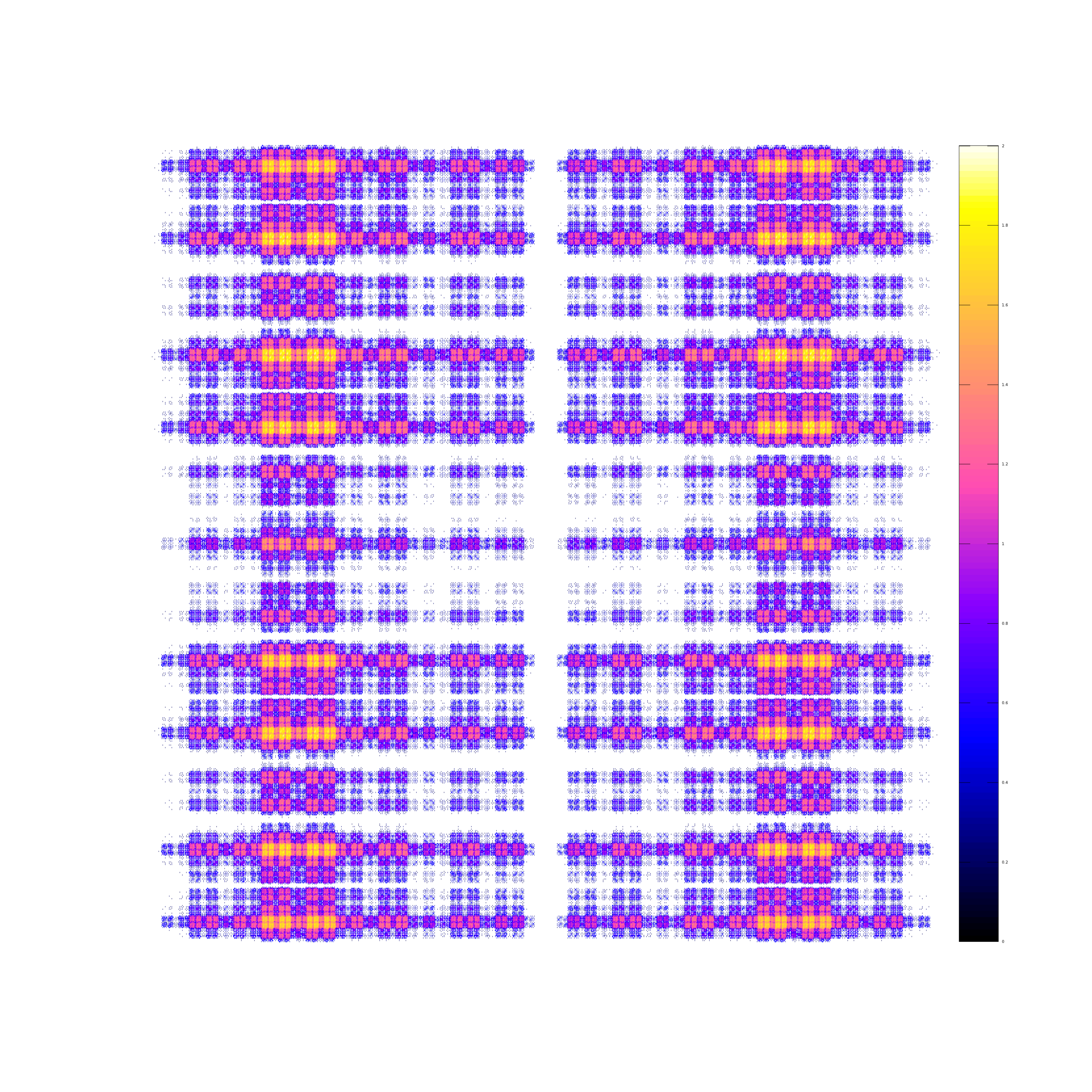}}\vspace{-0.1cm}
  \footnotesize\caption{Similarities of wave functions for the two different generalized labyrinth tilings $\mathcal{L}_a$ and $\mathcal{L}_a^\star$ shown for the silver mean model and for two different approximants of the golden mean model. The point size and color encode the probability $|\Phi|^2$ with a logarithmic scale, cp.~\cite{PhysRevB.1998.Repetowicz}.}
  \label{fig:wavefunctions}
\end{figure}

The characteristics of the wave functions $\Phi$ for the different tilings are interesting. Figure \ref{fig:wavefunctions} shows several wave functions of the generalized labyrinth tilings $\mathcal{L}_a$ and $\mathcal{L}_a^\star$ in two dimensions. For the silver mean model the wave functions are identical for $\mathcal{L}_a$ and $\mathcal{L}_a^\star$ if one of them is mirrored at the centre of the x- or y-axis, cp.~Fig.~\ref{fig:wavefunctions}(a) and \ref{fig:wavefunctions}(d). Both wave functions are also identical if rotated by 90$^\circ$ taking into account that the indices $i$ and $j$ of one of the eigenstates have to be interchanged. For other inflation rules the wave functions are not necessarily identical for small approximants due to the different boundaries of the lattice. Nevertheless, they are quite similarly distributed over the lattice as e.g.~displayed in Fig.~\ref{fig:wavefunctions}(b) and \ref{fig:wavefunctions}(e) and several of the wave functions show even a symmetry with respect to the x- or y-axis. For larger approximants the wave functions approach each other because the influence of the boundaries and of the phason flip for $b > 2$ vanish, see Fig.~\ref{fig:wavefunctions}(c) and \ref{fig:wavefunctions}(f). This characteristic can be helpful for analytical considerations for properties of the general labyrinth tiling.

\section{Scaling behavior of participation ratios}\label{sec:participation}

The participation ratio $p$ reflects the fraction of the total number of sites for which the probability measure of the wave function    $|\Phi_{\vek{r}}|^{2}$ is significantly different from zero and, hence, provides information about the degree of localization of the wave function $\Phi_{\vek{r}}$. For the discrete positions $\vek{r}$ (all valid combinations of the indices $l$, $m$, and $n$) it is defined via the inverse participation number $P$ and the number of states $V_a$ ($=N_a^d/2^{d-1}$ for even $N_a$) in $d$ dimensions as
\begin{equation}\label{equ:participation.1}
  p(\Phi,V_a) = \frac{P(\Phi)}{V_a} =   \frac{1}{V_a} \left[ \sum_{\vek{r}} |\Phi_{\vek{r}}|^{4} \right]^{-1}\;.
\end{equation}

The participation ratio $p(\Phi,V_a)$ of a state scales with a power law in the number of sites as $p(\Phi,V_a)
\sim V_a^{-\gamma(\Phi)}$,\index{scaling exponent!of participation ratio} where the exponent $\gamma$ is connected to the nature of the eigenstates \cite{Quasicrystals.2003.Grimm}. The wave function of a localized state is characterized by $\gamma = 1$, and $\gamma = 0$ corresponds to an extended state. For intermediate values of $\gamma$ ($0 < \gamma < 1$) we obtain fractal eigenstates, which are neither extended over the whole system nor completely localized at a certain position and show self-similar patterns \cite{PhysRevB.1998.Repetowicz, Thesis.2008.Thiem, PhysRevB.1987.Kohmoto}.

\begin{figure}[b]
 \centering
  \includegraphics[height=7.5cm]{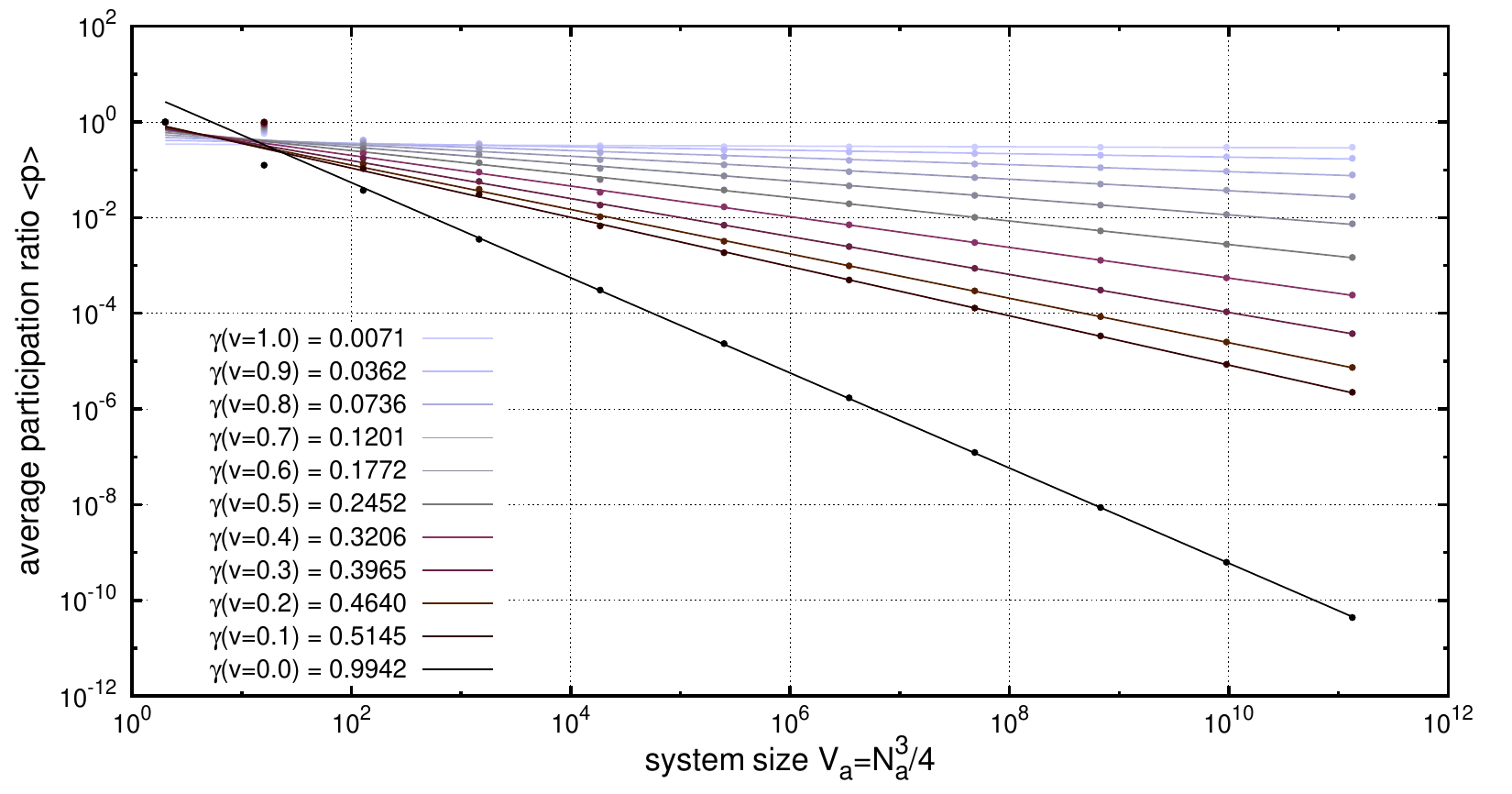}
  \caption{Scaling behavior of $\langle p \rangle$ for the silver mean model in $d=3$. The dots indicate $\langle p \rangle$ for different system sizes and the lines correspond to a mean least squares fit, where the slope equals the scaling exponent $\gamma$.}
  \label{fig:pscaling_oct}
\end{figure}

\begin{figure}
 \centering
  \subfigure[golden mean model]{\includegraphics[width=5.1cm]{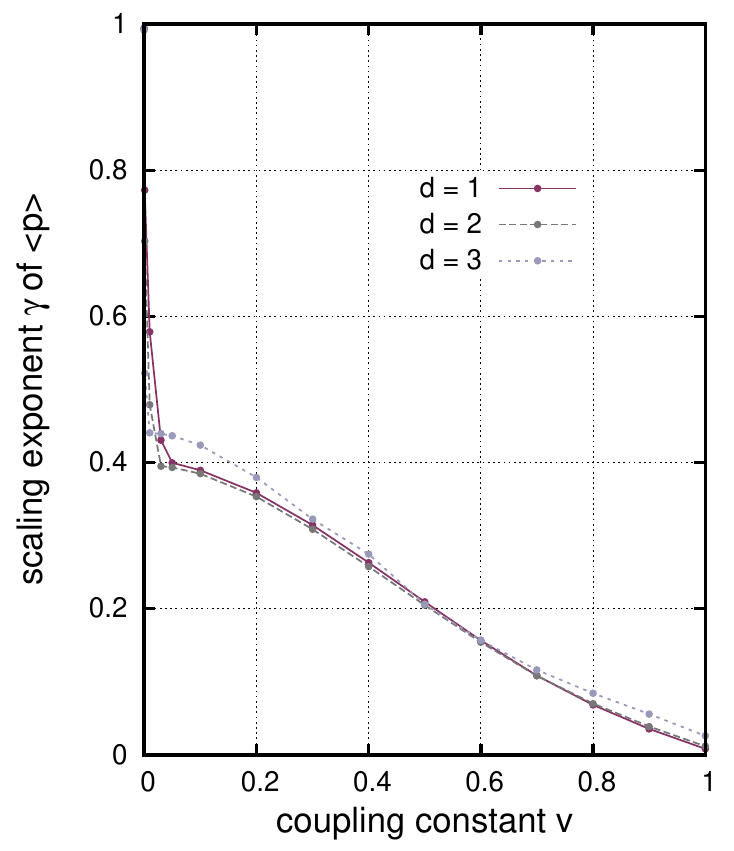}}\hspace{0.2cm}
  \subfigure[silver mean model]{\includegraphics[width=5.1cm]{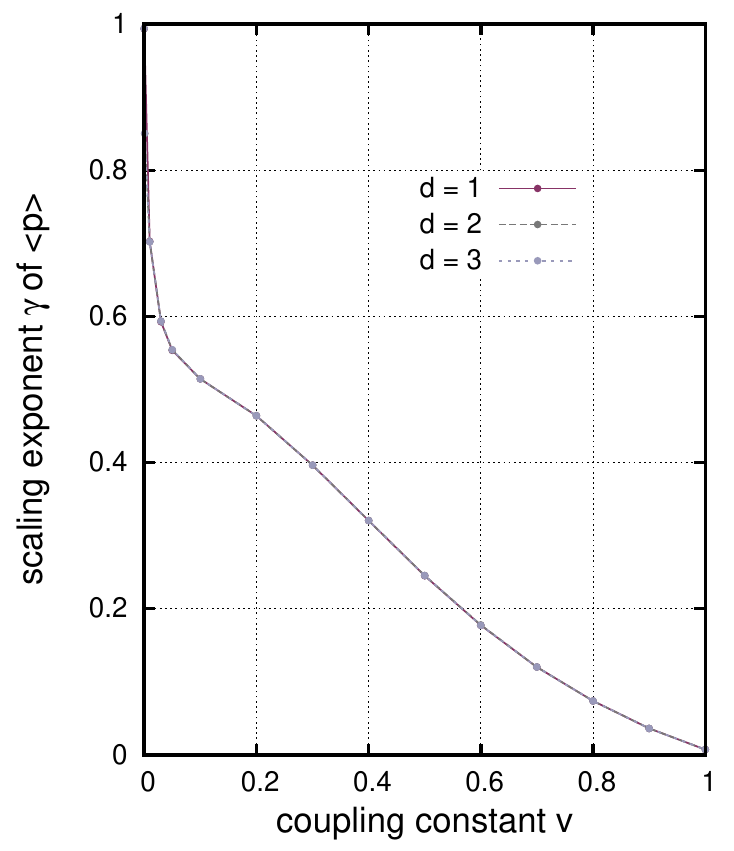}}\hspace{0.2cm}
  \subfigure[bronze mean model]{\includegraphics[width=5.1cm]{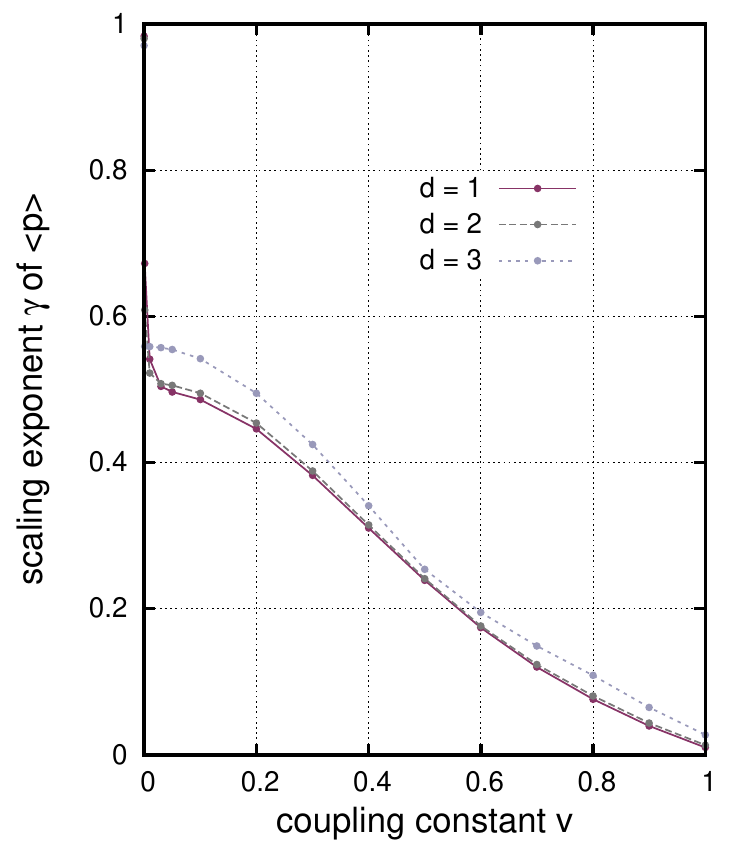}}
  \caption{Scaling exponents $\gamma$ of the average participation ratio $\langle p\rangle$ over the complete energy spectrum in
   one, two, and three dimensions for different metallic mean models.}
  \label{fig:pscaling_exp}
\end{figure}

Usually, the participation ratios in a small energy interval are distributed over a certain range resulting in a snow-flurry-like appearance when plotted. However, the average participation ratio is almost identical for states at the centre and the edge of the energy spectrum for the present model \cite{Thesis.2008.Thiem}. For other quasiperiodic tilings differences can be observed, e.g. the participation ratios of  eigenstates of the Penrose tiling, the generalized Rauzy tiling and octogonal tiling models are smaller at the centre than at the edges of the spectrum, whereas e.g. the Ammann-Kramer-Neri tiling shows an inverse behavior \cite{Quasicrystals.2003.Grimm, PhysRevB.1992.Passaro, PhysRevB.2002.Triozon}.
In order to compare the characteristics of the eigenstates for different dimensions we compute the scaling behavior for the average participation ratio $ \langle p(\Phi, V_a) \rangle = \frac{1}{V_a} \sum_{\vek{s}} p(\Phi^{\vek{s}},V_a)$. Results for the silver mean model for three dimensions are shown in Fig.~\ref{fig:pscaling_oct}. The average participation ratio scales very well with the system size. In Fig.~\ref{fig:pscaling_exp} we show the dependency of the scaling exponent $\gamma$ on the coupling constant $v$ for different metallic means in one, two, and three dimensions. The plot indicates that the scaling exponent $\gamma$ is independent of the dimension, which can be also analytically proven for the limit of infinite systems. The deviations for the different dimensions, especially for $d=3$, for the golden and bronze mean model are caused by the higher computational efforts allowing only the consideration of comparatively small systems. For the octonacci chain participation ratios of higher-dimensional tilings are given  as a product of the participation ratios of the chains \cite{PhysRevB.2005.Cerovski}.

\begin{figure}[h]
 \centering
  \subfigure[golden mean model ($b=1$)]{\includegraphics[height=6.7cm]{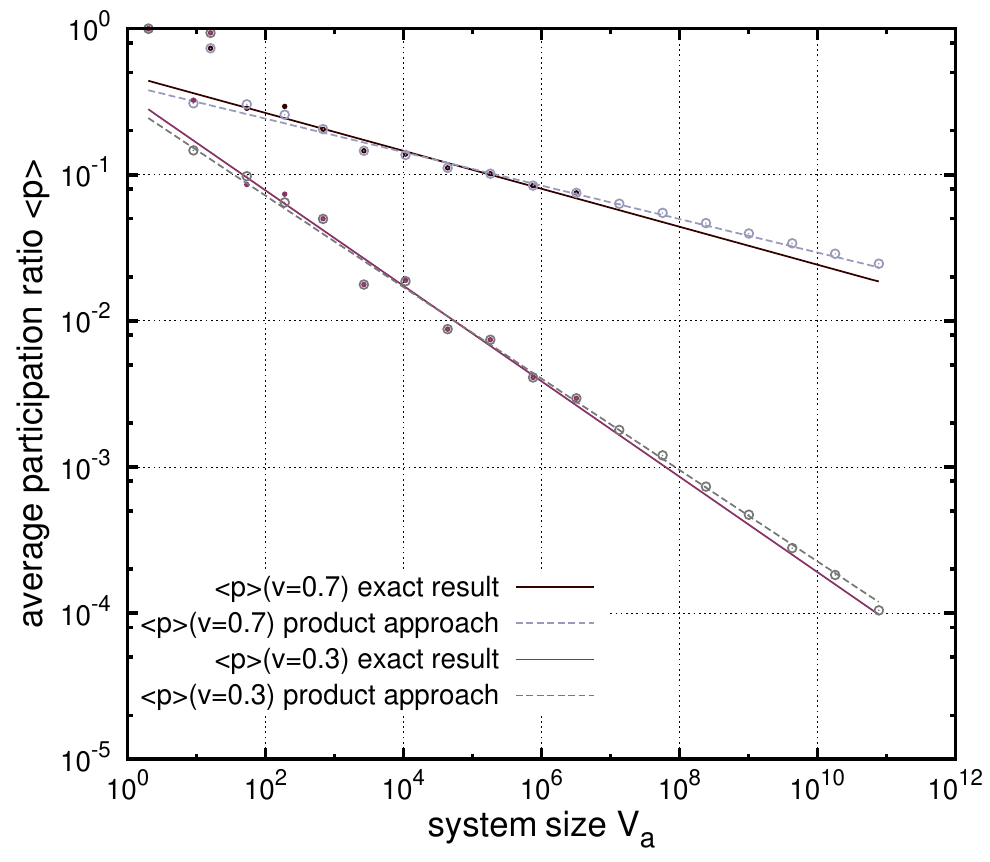}}\hspace{0.2cm}
  \subfigure[bronze mean model ($b=3$)]{\includegraphics[height=6.7cm]{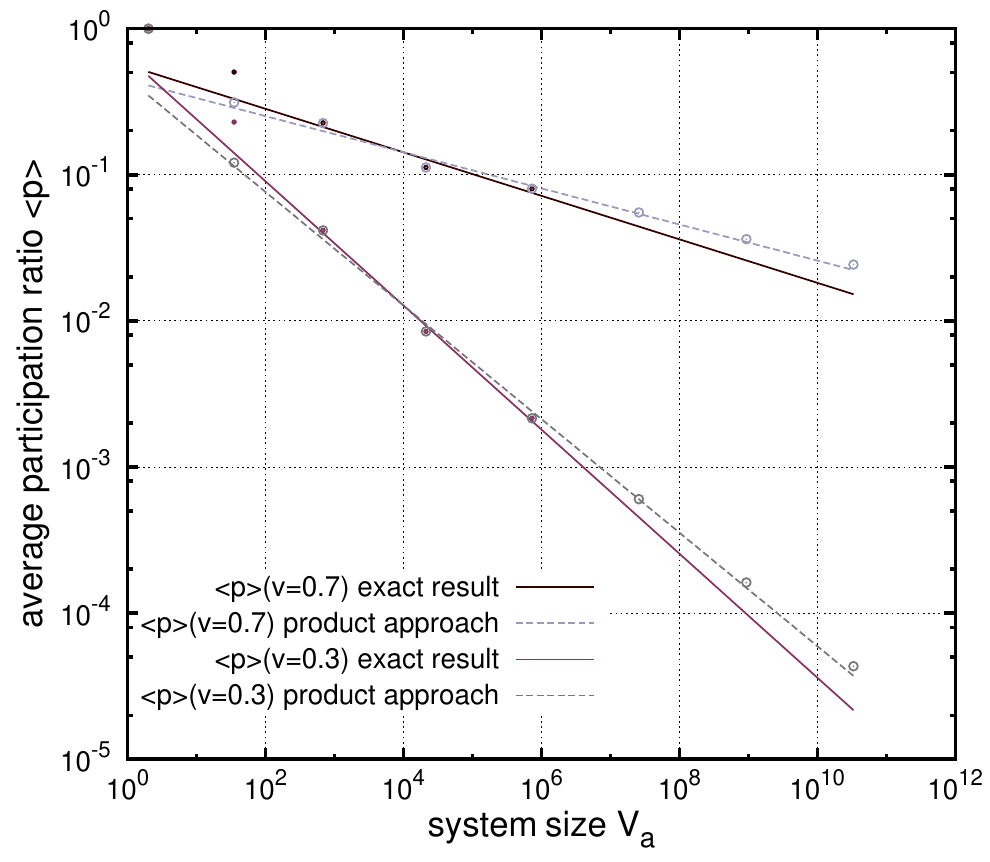}}
   \caption{Difference of the average participation ratio for the exact numerical result with the results obtained using a product approach for different generalized labyrinth tilings in three dimensions.}
  \label{fig:pscaling_approx}
\end{figure}

The similarities of the wave functions of the different general labyrinth tilings allow us to find good approximations for the calculation of certain properties such as participation ratios. In Fig.~\ref{fig:pscaling_approx} we compare the approximate results obtained by the product approach with the exact results and find that the deviations become very small already for system sizes with $N_a \approx 20$.

\section{Conclusion}\label{sec:conclusion}
In this paper we investigated the eigenstates for the generalized labyrinth tilings. We presented the construction rules for the eigenstates in two and three dimensions and found that for inflation rules with $b \ne 2$ the wave functions of the $2^{d-1}$ different tilings in $d$ dimensions approach each other with increasing system sizes in analogy to the occurrence of identical wave functions for $b=2$ for the different grids. This property is useful for approximate calculations of certain properties as shown for the participation ratios but also for analytical considerations.

\begin{ackn}
The authors like to thank the conference organizers for financial support to one of them (S. T.) by the International Union of Crystallography to attend the conference Aperiodic'09 and the Stiftung der Deutschen Wirtschaft for funding the research.
\end{ackn}

\section*{References}

\providecommand{\newblock}{}

\end{document}